\documentclass[a4paper,11pt]{article}
\usepackage{amsmath}
\usepackage{amssymb}

\DeclareMathOperator*{\BBcap}{\cap}

\def\dis{\displaystyle}

\def\BR{\relax\ifmmode {\mathbb R} \else${\mathbb R}$\fi}
\def\BC{\relax\ifmmode {\mathbb C} \else${mathbb C} $\fi}
\def\A{\relax\ifmmode {\mathcal A}\else${\mathcal A}$\fi}
\def\OA{\relax\ifmmode {\overline{\A}}\else${\overline{\A}}$\fi}
\def\AA{\relax\ifmmode {\mathfrak A}\else${\mathfrak A}$\fi}
\def\B{\relax\ifmmode {\mathcal B}\else${\mathcal B}$\fi}
\def\D{\relax\ifmmode {\mathcal D}\else${\mathcal D}$\fi}

\def\H{\relax\ifmmode {\mathcal H}\else${\mathcal H}$\fi}
\def\K{\relax\ifmmode {\mathcal K}\else${\mathcal K}$\fi}
\def\N{\relax\ifmmode {\mathcal N}\else${\mathcal N}$\fi}

\def\sharp{\text{{\scriptsize $\#$}}}

\begin{document}
\begin{center}
{\bf \Large Some classes of topological quasi $*$-algebras}

\ 

{\large 
F. Bagarello$^{1}$, A. Inoue$^{2}$ and C. Trapani$^{3}$ } \vspace{3mm} \\ 

$^{1}$ Dipartimento di Matematica dell'Universit\`{a} di Palermo\\ 
I-90128 Palermo, Italy \vspace{2mm}\\
$^{2}$ Department of Applied Mathematics \\Fukuoka University\\
J-814-80 Fukuoka, Japan \vspace{2mm}\\
$^{3}$ Dipartimento di Scienze Fisiche ed Astronomiche\\ dell' Universit\`{a} di Palermo\\  I-90123 Palermo, Italy
\end{center}

\vspace{1cm}

\ 

{\bf Abstract.}
The completion $\overline{\A}[\tau]$ of a locally convex $*$-algebra $\A [ \tau ]$ with not jointly continuous multiplication is a $*$-vector space with partial multiplication $xy$ defined only for $x$ or $y \in \A_{0}$, and it is called a topological quasi $*$-algebra.
In this paper two classes of topological quasi $*$-algebras called strict CQ$^*$-algebras and HCQ$^*$-algebras are studied. Roughly speaking, a strict CQ$^*$-algebra (resp. HCQ$^*$-algebra) is a Banach (resp. Hilbert) quasi $*$-algebra containing a C$^*$-algebra endowed with another involution $\sharp$ and C$^*$-norm $\| \ \|_{\sharp}$. HCQ$^*$-algebras are closely related to left Hilbert algebras. We shall show that a Hilbert space is a HCQ$^*$-algebra if and only if it contains a left Hilbert algebra with unit as a dense subspace. Further, we shall give a necessary and sufficient condition under which a strict CQ$^*$-algebra is embedded in a HCQ$^*$-algebra.

\ 

\section{Introduction}
Topological quasi $*$-algebras were first introduced by Lassner \cite{lassner1} for the mathematical description of some quantum physical models, and after that, they have been studied by Lassner \cite{lassner1, lassner2}, Trapani \cite{trapani} and Bagarello-Trapani \cite{bt3,bt4} etc. In this paper we shall study two classes of topological quasi $*$-algebras called strict CQ$^*$-algebras and HCQ$^*$-algebras from a mathematical point of view but also in the perspective of possible physical applications. Let $\A$ be a $*$-algebra with two involutions $*$ and $\sharp$ and two norms $\| \ \|$ and $\| \ \|_{\sharp}$ satisfying $\| x^* \| = \| x \|, \| x \| \leq \| x \|_{\sharp}$ and $\| x^{\sharp} x\|_{\sharp} = \| x \|^2_{\sharp}$ for each $x,y \in \A$. 
Then the completion $\OA[\| \ \|]$ of $\A [\| \ \|]$ is a topological quasi $*$-algebras containing (under natural assumptions) two C$^*$-algebras $\A_{\sharp}[\| \ \|_{\sharp} ]$ and $\A_{\flat}[\| \ \|_\flat ]$ with different involutions $\sharp$ and $\flat$, respectively, which are connected by the isometric involution $J: x \rightarrow x^*$. This is called a {\it pseudo CQ$^*$-algebra}. If $\| x \|_{\sharp} = \sup\{ \| xy \| \ ; \ \| y\| \leq 1 \}$, then $\OA[\| \ \|]$ is a particular kind of CQ$^*$-algebra as  defined and studied in \cite{bt1,bt2}, and it is called a {\it strict CQ$^*$-algebra}, and denoted by $(\OA[\| \ \|], \sharp, \| \ \|_{\sharp})$. Let $\OA[ \| \ \| ]$ be a topological quasi $*$-algebra with isometric involution $J: x \rightarrow x^*$ and Hilbertian norm $\| \ \|$. 
If $\A$ has another involution $\sharp$ satisfying $\| x \| \leq \| L_x \|$ and $L_x^* = L_{x^{\sharp}}$ for each $x \in \A$, where $L_x$ is the bounded linear operator on the Hilbert space $\OA[\| \ \|]$ defined by $L_x y = xy, y \in \A$, then $\OA[ \| \ \|]$ is a strict CQ$^*$-algebra with involution $\sharp$ and C$^*$-norm $\| x \|_{\sharp} \equiv \| L_x \|, x \in \A$, and it is called a {\it HCQ$^*$-algebra} and denoted by ($\OA[\| \ \|], \sharp)$. HCQ$^*$-algebras are closely related to left Hilbert algebras. Let $(\OA[\| \ \|],\sharp)$ be a HCQ$^*$algebra. Then $\A$ is a left Hilbert algebra in the Hilbert space $\OA[\| \ \|]$ with involution $\sharp$, and the full left Hilbert algebra $\A''$ of $\A$ has unit. But, the isometric involution $J$ does not necessarily coincide with the modular conjugation operator $J_\A$ of the left Hilbert algebra $\A$. If $J_\A = J$, then the HCQ$^*$-algebra $(\OA[\| \ \|], \sharp)$ is said to be {\it standard}.
Suppose that $(\OA[\| \ \|], \sharp)$ is standard. Then $\A$ is contained in the maximal Tomita algebra $(\A'')_0$ of $\A''$ and $ (\overline{(\A'')_0}[ \| \ \|], \sharp)$ is a standard HCQ$^*$-algebra with the one-parameter group $\{ \Delta^{it}_\A \}_{t \in \BR}$ of $*$-automorphisms, where $\Delta_\A$ is the modular operator of $\A$. >From these results, it is shown that a Hilbert space is a standard HCQ$^*$-algebra if and only if it contains a left Hilbert algebra as dense subspace. Finally we give a necessary and sufficient condition under which a strict CQ$^*$-algebra is embedded into a standard HCQ$^*$-algebra using the GNS-construction of positive sesquilinear form on the strict CQ$^*$-algebra $\OA[\| \ \|]$.

\section{Strict CQ$^*$-algebras and HCQ$^*$-algebras}
Let $\A[\| \ \|]$ be a normed $*$-algebra with isometric involution $*$ and separately (but not jointly) continuous multiplication. Then the completion, $\OA[ \| \ \|]$, of $\A[\| \ \|]$ is a topological quasi $*$-algebra that we call, as is natural, a {\it Banach quasi $*$-algebra}. In particular, if $\| \ \|$ is a Hilbertian norm, then $\OA[ \| \ \|]$ is called a {\it Hilbert quasi $*$-algebra}. 
For any $a \in \OA[ \| \ \|]$ we put 
\[
L_a x = ax \text{ and } R_a x = xa, \hspace*{5mm} x \in \A.
\]
Then $L_a$ and $R_a$ are linear maps of $\A$ into $\OA[\| \ \|]$. 
In particular, if $a \in A$, then $L_a$ and $R_a$ can be extended to bounded linear operators on the Banach space $\OA[\| \ \|]$ and they are denoted by the same symbols $L_a$ and $R_a$.

Let $\OA[ \| \ \|]$ be a Banach quasi $*$-algebra and assume that the $*$-algebra $\A$ has another norm $\| \ \|_{\sharp}$ and another involution $\sharp$ satisfying the following conditions:
\\
\hspace*{6mm}
\begin{minipage}[t]{12cm}
\begin{itemize}
\item[(a.1)]$ \| x^{\sharp}x\|_{\sharp}= \|x\|_{\sharp}^2, \quad \forall x \in \A$.
\item[(a.2)]$ \|x\| \leq \|x\|_{\sharp}, \quad \forall x \in \A$.
\item[(a.3)]$ \|xy\| \leq \|x\|_{\sharp}\|y\|, \quad \forall x,y \in \A$.
\end{itemize}
\end{minipage}\\[3mm]
Then by (a.2), the identity map $i: \A[\| \ \|_{\sharp}] \to \A [\| \ \|]$ has a continuous extension $\hat{i}$ from the completion $\A_{\sharp}$ of $\A[\| \ \|_{\sharp}]$ ($\A_{\sharp}$ is, of course, a C*-algebra) into $\OA[\| \ \|]$. If $\hat{i}$ is injective, then $\A_{\sharp}$ is (identified with) a dense subspace of $\OA$. This happens if, and only if,\\ 
\hspace*{6mm}
\begin{minipage}[t]{12cm}
\begin{itemize}
\item[(a.4)] two norms $\| \ \|$ and $\| \ \|_{\sharp}$ are compatible in the following sence \cite{gelfand}: for any sequence $\{x_n\}\subset \A$ such that $\|x_n\|\to 0$ and $x_n \to x$ in $\A_{\sharp}[\| \ \|_{\sharp}]$, $x=0$ results, i.e. if $\,\hat{i}^{-1}: \A [\| \ \|] \to \A_{\sharp}[\| \ \|_{\sharp}]$ is closable.

\end{itemize}
\end{minipage}\\
 
\ 

{\bf Definition 2.1.} A Banach quasi $*$-algebra $\OA[ \| \ \|]$ is said to be a {\it psuedo CQ$^*$-algebra} if the $*$-algebra $\A$ has a
another norm $\| \ \|_{\sharp}$ and another involution $\sharp$ satisfying the conditions (a.1)-(a.4) above. 
Further, if $\| x \|_{\sharp} = \| L_x \| \equiv \sup \{ \| xy \|; y \in \A
\text{ s.t. } \| y \| \leq 1 \}$ for each $x \in \A$, then $\OA[\| \ \|]$ is said to be a {\it strict CQ$^*$-algebra}.

\ 

A pseudo CQ$^*$-algebra  $\OA[\| \ \|]$ is fully determined by the involution $\sharp$ and the C$^*$-norm $\| \ \|_{\sharp}$, and so it will be often denoted by $(\OA[\| \ \|], \sharp, \| \ \|_{\sharp})$. On the other hand, a strict CQ$^*$-algebra is fully determined when the new involution $\sharp$ is known; so it can be simply denoted as $(\OA[\| \ \|], \sharp)$, making lighter in this way the notation introduced by two of us in \cite{bt1,bt2}. 
Let $(\OA[\| \ \|], \sharp, \| \ \|_{\sharp})$ be a pseudo CQ$^*$-algebra and, as above, $\A_{\sharp}$ the C$^*$-algebra obtained by completing the $\sharp$-algebra $\A$ with respect to the C$^*$-norm $\| \ \|_{\sharp}$. Let $J$ be the involution $*$ of the Banach quasi $*$-algebra $\OA[\| \ \|]$. Then $\A_\flat \equiv J \A_{\sharp}$ is a C$^*$-algebra equipped with the operations:
$x^* + y^* \equiv (x+y)^*, \lambda x^* \equiv (\overline{\lambda}x)^*, x^* y^* \equiv (yx)^*$, the involution $(x^*)^\flat \equiv x^{\sharp *}$ and the C$^*$-norm $\| x^* \|_\flat \equiv \| x \|_{\sharp}, \forall x,y\in\A_\flat$. 

\ 

{\bf Proposition 2.2.} A pseudo CQ$^*$-algebra $(\OA[ \| \ \|], \sharp, \| \ \|_{\sharp})$ contains two C$^*$-algebras $\A_{\sharp}$ and $\A_\flat \equiv J \A_{\sharp}$ with different involutions $\sharp$ and $\flat$, respectively, as dense subalgebra. In particular, if $(\OA[\| \ \|], \sharp, \| \ \|_{\sharp})$ is a strict CQ$^*$-algebra, then $L_{\A_{\sharp}}$ and $R_{\A_{\flat}}$ are C$^*$-algebras, $L_x R_y = R_y L_x$ for each $x \in \A_{\sharp}$ and $y \in \A_\flat$ and $R_{\A_\flat} = J L_{\A_{\sharp}} J$.

\ 

By Proposition 2.2, every strict CQ$^*$-algebra is a CQ$^*$-algebra in the sense of \cite{bt1,bt2} but the converse is not true in general. \\
We summarize the situation with the following scheme
\[ \begin{array}{ccccc}
                 & \subset & \A_{\sharp}                       & \subset & \\
\A[\| \ \|]      &         &  \hspace*{3mm} \updownarrow J &         & \OA[\| \ \|] \\
                 & \subset & \A_\flat                      & \subset & \\
\text{normed $*$-algebra} &         & \text{C$^*$-algebras}          & &\text{ CQ$^*$-algebra},
\end{array}\]
which summarizes the situation: the *-algebra $\A[\| \ \|]$ is contained in  its closures,  $\A_{\sharp}=\OA[\| \ \|_{\sharp}]$ and $\A_{\flat}=\OA[\| \ \|_\flat]=J\A_{\sharp}$. These C$^*$-algebras, moreover, are both contained in $\OA[\| \ \|]$. 

\ 

{\bf Definition 2.3.} A Hilbert quasi $*$-algebra $\OA[\| \ \|]$ is said to be a {\it HCQ$^*$-algebra} if there is another involution $\sharp$ of $\A$ such that $L^*_x = L_{x^{\sharp}}$ and $\| x \| \leq \| L_x \|$ for each $x \in \A$. 
Here we denote it by $(\OA[\| \ \|], \sharp)$.

\ 

HCQ$^*$-algebras are closely related to left Hilbert algebras. Before going forth, for reader's convenience, we breafly review the definitions and the basic properties about left Hilbert algebras. 
A $*$-algebra $\AA$ with involution $\sharp$ is said to be a {\it left Hilbert algebra} if it is a dense subspace in a Hilbert space $\H$ with inner product $( \ | \ )$ satisfying the following conditions:

(i) For any $x \in \AA$ the map $y \in \AA \rightarrow xy \in \AA$ is continuous. 

(ii) $(xy|z) = (y|x^{\sharp}z), \hspace*{5mm} {}^\forall x, y, z \in \AA$.

(iii) $\AA^2 \equiv \{ xy ; x,y \in \A \} \text{ is total in } \H$.

(iv) The involution $x \rightarrow x^{\sharp}$ is closable in $\H$.\\
By (i), for any $x \in \AA$ we denote by $\pi_{\AA}(x)$ the unique continuous linear extension to $\H$ of the map $y \in \AA \to xy \in \AA$; then $\pi_\AA$ is a $*$-representation of $\AA$ on $\H$. We denote by $S_{\AA}$ the closure of the involution $\sharp$. Let $S_\AA = J_\AA \Delta^{\frac{1}{2}}_\AA$ be the polar decomposition of $S_\AA$. Then $J_\AA$ is an isometric involution on $\H$ and $\Delta_\AA$ is a non-singular positive self-adjoint operator in $\H$ such that $S_\AA= J_\AA \Delta^{\frac{1}{2}} _\AA= \Delta^{-\frac{1}{2}}_\AA J_\AA$ and $S^*_\AA= J_\AA \Delta^{-\frac{1}{2}}_\AA= \Delta^{\frac{1}{2}}_\AA J_\AA$, and $J_\AA$ is called the {\it modular conjugation operator} of $\AA$ and $\Delta_\AA$ is called the {\it modular operator} of $\AA$. We define the commutant $\AA'$ of $\AA$ as follows: For any $y \in \D(S^*_\AA)$ we put $\pi'_\AA(y) x = \pi_\AA(x) y, x \in \AA$ and put $\AA'= \{ y \in \D(S^*_\AA); \pi'_\AA(y) \text{ is bounded } \}$. Then $\AA'$ is a left Hilbert algebra in $\H$ with involution $S^*_\AA$ and multiplication $y_1 y_2 \equiv \pi'_\AA(y_2)y_1$. Similarly, the commutant $\AA''$ of $\AA'$ is defined by $\AA''= \{ x \in \D(S_\AA); y \in \AA' \rightarrow xy \text{ is continuous } \}$. For any $x \in \AA''$ we denote by $\pi_\AA(x)$ the unique continuous linear operator on $\H$ such that $\pi_\AA(x) y= \pi'_\AA(y) x, y \in \AA'$. Then $\AA''$ is a left Hilbert algebra in $\H$ with involution $S_\AA$ and multiplication $x_1 x_2 \equiv \pi_\AA(x_1)x_2$ containing $\AA$. A left Hilbert algebra $\AA$ is said to be {\it full} if $\AA= \AA''$. It is well-known as the Tomita fundamental theorem that $J_\AA \pi_\AA(\AA)'' J_\AA= \pi_\AA (\AA)'$ and $\Delta^{it}_\AA \pi_\AA(\AA)'' \Delta^{-it}_\AA = \pi_\AA(\AA)'', {}^\forall t \in \BR$. 
Let $\AA$ be a full left Hilbert algebra in $\H$, and $\AA_0 \equiv \{ x \in \dis \BBcap_{\alpha \in \BC} \D(\Delta^\alpha_\AA); \Delta^\alpha_\AA x \in \AA, {}^\forall \alpha \in \BC \}$. 
Then $\AA_0$ is a left Hilbert subalgebra in $\H$ such that $\AA''_0= \AA, J_\AA \AA_0 = \AA_0$ and $\{ \Delta^\alpha_\AA ; \alpha \in \BC \}$ is a complex one-parameter group of automorphisms of $\AA_0$ such that $(\Delta^\alpha_\AA x)^{\sharp}= \Delta^{-\overline{\alpha}}_\AA x^{\sharp}$ and $(\Delta^\alpha_\AA x )^*= \Delta^{-\overline{\alpha}}_\AA x^*$ for each $\alpha \in \BC$ and $x \in \AA_0$. 
This $\AA_0$ is called the {\it maximal Tomita algebra} of $\AA$.
For more details, we refer to \cite{stsz,takesaki,van}.

\

{\bf Proposition 2.4.} Suppose that $(\OA[\| \ \|], \sharp)$ is a HCQ$^*$-algebra. 
Then the following statements hold:

(i) $(\OA[ \| \ \|], \sharp)$ is a strict CQ$^*$-algebra with the C$^*$-norm $\| x \|_{\sharp} = \| L_x \|, x \in \A$.

(ii) $\A$ is a left Hilbert algebra in the Hilbert space $\H \equiv \OA[\| \ \|]$ whose full left Hilbert algebra $\A''$ has a unit $u$. \\[3mm]
\indent
{\it Proof.}
(i) The proof is mostly trivial. We prove only that the condition (a.4) is satisfied in this case. Indeed, if $\{x_n\}\subset \A$ is a sequence such that $\|x_n\|\to 0$ and $x_n \to x$ in $\A_{\sharp}[\| \ \|_{\sharp}]$, then by the assumption $L_{x_n} \to L_x$ with respect to the operator norm. The continuity of the multiplication in $\A[\| \ \|]$ easily implies that $L_x=0$; thus $\|x\|_{\sharp}=0$ and $x=0$. \\
(ii) We first show that $\A$ is a left Hilbert algebra in $\H$ with involution $\sharp$. Since the C$^*$-algebra $\A_{\sharp}$ has an approximate identity $\{ u_\alpha \}, \A$ is dense in the C$^*$-algebra $\A_{\sharp}$ and $\| x \| \leq \|x \|_{\sharp}$ for each $x \in \A$, and then it follows that $\A^2$ is total in $\A[\| \ \|]$. The assumption $L_x^* = L_{x^{\sharp}} ({}^\forall x \in \A)$ implies that $(xy|z) = (y | x^{\sharp}z)$ for each $x,y,z \in \A$, where $(\ | \ )$ is the inner product defined by the Hilbertian norm $\| \ \|$. Further, we have $\pi_\A(x) = L_x, {}^\forall x \in \A$ and $\pi_\A(x)$ is bounded. Take any sequence $\{ x_n \} $ in $\A$ such that $\dis \lim_{n \rightarrow \infty} \| x_n \|=0$ and $\dis \lim_{n \rightarrow \infty} \| {x_n}^{\sharp}-y \|=0$. 
Then it follows that $(y | x_1 x_2^\flat) = \dis \lim_{n \rightarrow \infty} (x_n^{\sharp} | x_1 x_2^\flat) =\dis \lim_{n \rightarrow \infty} ( x_2 x_1^\flat| x_n) = 0$ for each $x_1, x_2 \in \A$, which implies that $x \in \A \mapsto x^{\sharp} \in \A$ is closable. 
Thus $\A$ is a left Hilbert algebra in $\H$ with the involution $\sharp$. 
We next show that the full left Hilbert algebra $\A''$ has a unit $u$. 
For any $\varepsilon > 0$ and for each finite subsets $\{ x_1, \dotsc, x_m \}$ and $\{ y_1, \dotsc, y_m\}$ of $\A$, we define the set
\begin{align*}
&K(\varepsilon; \{ x_1, \dotsc, x_m \}, \{ y_1, \dotsc, y_m \})\\
&\hspace*{8mm} = \{ a \in \H ; \| a \| \leq 1, | (a x_k -x_k | y_k) | \leq \varepsilon \\
&\hspace*{28mm} \text{ and } | (x_k a-x_k| y_k) | \leq \varepsilon, k = 1, \dotsc, m \}.
\end{align*}
Since the C$^*$-algebra $\A_{\sharp}$ has an approximate identity and $\|x\| \leq \| x \|_{\sharp}$ for each $x \in \A_{\sharp}$, it follows that $K(\varepsilon; \{ x_1, \dotsc, x_m \}, \{ y_1, \dotsc, y_m \}) \neq \phi$. 
Let now $\K$ be the family of all subsets $K(\varepsilon; \{ x_1, \dotsc, x_m \}, \{ y_1, \dotsc, y_m \})$ where $\varepsilon > 0$ and $\{ x_1, \dotsc, x_m \}, \{ y_1, \dotsc, y_m \}$ are finite subsets. 
Then $\K$ is a family of non-empty weakly closed subsets of the weakly compact set $\H_1 \equiv \{ a \in \H; \| a \| \leq 1 \}$. 
Hence the intersection of all the sets in $\K$ is non empty. 
Hence, an element  $u$ of this intersection is such that $u$ is a {\it quasi-unit} of the topological quasi $*$-algebra $\OA[\| \ \|]$ , that is, $u \in \OA[ \| \ \|]$ and $ux = xu = x$ for each $x \in \A$. 
Since
\[
(S_\A x | u) = (x^{\sharp} |u) = (u |L_x u) = (u|x) 
\]
for each $x \in \A$, it follows that $u \in \D(S_\A^*)$ and $\pi'_\A(u)=I$. 
Hence, $u \in \A'$ and $S_\A^*u = u$, which implies that 
\[
(S_\A^* y| u) = ( \pi'_\A (S^*_\A y)u | u) = (u | \pi'_\A(y)u) = (u|y)
\]
for each $y \in \A'$. 
Hence we have $u \in \A''$ and $S_\A u = u$. 
This completes the proof. 

\ 

By Proposition 2.4, the situation of HCQ*-algebras can be schematized with the following diagram.
\[
\begin{array}{ccccccccc}
  & \subset & \A_{\sharp} & \subset & {\A}''= L_{\A}'' u & \subset & \D(S_\A) & \subset &  \\[3mm]
\A&       & \hspace*{-3mm} \updownarrow J &  & \hspace*{-3mm} \updownarrow J_\A & & \hspace*{-3mm} \updownarrow J_\A & & \OA[\| \ \|].\\[3mm]
  &\subset& \A_\flat &\subset& \A'={L_\A}' u& \subset& \D(S^*_{\A})& \subset& 
\end{array}
\]
We now look for conditions under which $J= J_\A$. 

\ 

{\bf Lemma 2.5.} Let $(\OA[ \| \ \|], \sharp)$ be a HCQ$^*$-algebra.
Then the following statements are equivalent:

(i) $J= J_\A$. 

(ii) $(x^{\sharp} | x^*) \geq 0$ for each $x \in \A$. \\[3mm]
\indent
{\it Proof.} (i) $\Rightarrow $ (ii) This follows from
\[
(x^{\sharp} | x^*) = (J_\A \Delta^{\frac{1}{2}}_\A x | J_\A x ) = (x | \Delta^{\frac{1} {2}}_\A x ) \geq 0, {}^\forall x \in \A.
\]
(ii) $\Rightarrow$ (i) By the assumption (ii) we have $S_\A= J(JJ_\A \Delta^{\frac{1}{2}}_\A)$ and $JJ_\A \Delta^{\frac{1}{2}}_\A \geq 0$. The uniqueness of the polar decomposition of $S_\A$ implies $J=J_\A$. 

\ 

If anyone of the two equivalent statements of Lemma 2.5 holds, we say that the HCQ$^*$-algebra $(\OA[ \| \ \|], \sharp)$ is {\it standard}.

\ 

{\bf Remark 2.6.} Let $(\OA[\| \ \|], \sharp)$ be a HCQ$^*$-algebra. If it is standard, then $R'_\A = L''_\A$. Conversely, if $R'_\A = L''_\A$, then $JJ_\A= J_AJ$, but we don't know whether $J= J_\A$. 

\ 

Since two HCQ$^*$-algebras $(\OA[\| \ \|], \sharp)$ with $(\overline{\B} [ \| \ \|], \sharp), \OA[\| \ \|] = \overline{\B}[\| \ \|]$ as Hilbert spaces, need not coincide as HCQ$^*$-algebras, we introduce the following notion:

\ 

{\bf Definition 2.7.} A HCQ$^*$-algebra $\OA[\| \ \|]$ is said to be an {\it extension} of a HCQ$^*$-algebra $\overline{\B}[ \| \ \|]$ if $\B$ is a dense $*$-subalgebra of $\A$ and $S_\A = S_\B$. 

\ 

{\bf Proposition 2.8.} Let $(\OA[\| \ \|], \sharp)$ be a standard HCQ$^*$-algebra, and $\B \equiv( \A'')_0$ the maximal Tomita algebra of the full left Hilbert algebra $\A''$. Then $(\overline{\B}[ \| \ \|], S_\A)$ is a standard HCQ$^*$-algebra and it is an extension of $(\OA[\| \ \|], S_\A)$. Further, $\{ \Delta^{it}_\A \}_{t \in \BR}$ is a one-parameter group of $*$-automorphisms of the Hilbert quasi $*$-algebra $\overline{\B}[ \| \ \|]$, that is, $\Delta^{it}_\AA \B = \B$, $(\Delta^{it}_\A a)^* = \Delta^{it}_\A a^*$, $\Delta^{it}_\A(ax) = (\Delta^{it}_\A a)(\Delta^{it}_\A x)$ and $\Delta^{it}_\A(xa) = (\Delta^{it}_\A x) (\Delta^{it}_\A a)$ for all $a \in \overline{\B}[\| \ \|], x \in \B$ and $t \in \BR$. \\[3mm]\indent
{\it Proof.} It is almost clear that $\overline{\B}(\| \ \|)$ is a Hilbert quasi $*$-algebra with the involution $J_\A = J_\B$ and further $(\overline{\B} [\| \ \|], S_\A)$ is a standard HCQ$^*$-algebra. 
Since $\{ \Delta^{it}_\A \}_{t \in \BR}$ is a one-parameter group of $*$-automorphisms of the Tomita algebra $\B$, it follows that $\{ \Delta^{it}_\A \} _{t \in \BR}$ is also a one-parameter group of $*$-automorphisms of the Hilbert quasi $*$-algebra $\overline{\B}[\| \ \|]$. 

\ 

Finally we consider the question of when a Hilbert space can be regarded as a standard HCQ$^*$-algebra. By Proposition 2.4, 2.8 and (\cite{takesaki} Theorem 13.1) we have the following

\ 

{\bf Theorem 2.9.} Let $\H$ be a Hilbert space. The following statements are equivalent:

(i) $\H$ is a standard HCQ$^*$-algebra.

(ii) $\H$ contains a left Hilbert algebra with unit as dense subspace.

(iii) There exists a von Neumann algebra on $\H$ with a cyclic and separating vector.

It is worth noticing, in particular, that the implication (iii) $\Rightarrow$ (i) shows that the class of standard HCQ$^*$-algebras is rather rich. 

\section{The structure of strict CQ$^*$-algebras}
In this section we study when a strict CQ$^*$-algebra is embedded in a standard HCQ$^*$-algebra. For that, we need a GNS-like construction for a class of positive sesquilinear forms on strict CQ$^*$-algebras $(\OA[\| \ \|], \sharp, \| \ \|_{\sharp})$. A sesquilinear form $\varphi$ on $\OA[\| \ \|] \times \OA[\| \ \|]$ is said to be {\it positive} if $\varphi(a,a) \geq 0$ for all $a \in \OA[\| \ \|]$, and $\varphi$ is said to be {\it faithful} if $\varphi(a,a)=0, a \in \OA[\| \ \|]$, implies $a = 0$. 
Further, we need the following notion:

\ 

{\bf Definition 3.1.} Let $(\OA[\| \ \|], \sharp, \| \ \|_{\sharp})$ and $(\overline{\B}[ \| \ \|_1], \sharp_1, \| \ \|_{\sharp_1})$ be strict CQ$^*$-algebras. 
A linear map $\Phi: \OA[\| \ \|] \rightarrow \overline{\B} [ \| \ \|_1]$ is said to be a {\it $*$-homomorphism} of $(\OA[\| \ \|], \sharp, \| \ \|_{\sharp})$ into $(\overline{\B}[ \| \ \|_1], \sharp_1, \| \ \|_{\sharp_1})$ if (i) $\Phi$ is a $*$-homomorphism of the quasi $*$-algebra $\OA[\| \ \|]$ into the quasi $*$-algebra $\overline{\B}[\| \ \|_1]$, that is, $\Phi(\A) \subset \B$ and $\Phi(a)^* = \Phi(a^*), \Phi(ax) =\Phi(a) \Phi(x)$ and $\Phi(xa)= \Phi(x) \Phi(a)$ for all $a \in \OA[\| \ \|]$ and $x \in \A$; 
(ii) $\Phi \lceil \A_{\sharp}$ is a $*$-homomorphism of the C$^*$-algebra $\A_{\sharp}$ into the C$^*$-algebra $\B_{\sharp_1}$. A bijective (resp. injective) $*$-homomorphism $\Phi$ such that $\Phi(\A) = \B$ and $\Phi(\A_{\sharp}) = \B_{\sharp_1}$ is called a {\it $*$-isomorphism} of $(\OA[ \| \ \|], \sharp, \| \ \|_{\sharp})$ onto (resp. into) $(\overline{\B}[\| \ \|_1], \sharp_1, \| \ \|_{\sharp_1})$. 
A $*$-homomorphism $\Phi$ is said to be {\it contractive} if $\| \Phi(a) \|_1 \leq \| a \|$ for all $a \in \OA[\| \ \|]$. A contractive $*$-isomorphism whose inverse is also contractive is called an {\it isometric $*$-isomorphism}. 

\ 

{\bf Theorem 3.2.} Let $(\OA[\| \ \|], \sharp, \| \ \|_{\sharp})$ be a strict CQ$^*$-algebra with quasi-unit $u$. Then the following statements are equivalent:

(i) There exists a contractive $*$-homomorphism (resp. $*$-isomorphism) of the strict CQ$^*$-algebra $(\OA[\| \ \|], \sharp, \| \ \|_{\sharp})$ into a HCQ$^*$-algebra $(\overline{\B}[\| \ \|_1], \sharp_1)$. 

(ii) There exists a (resp. faithful) positive sesquilinear form $\varphi$ on $\OA[\| \ \|] \times \OA[\| \ \|]$ satisfying \hspace*{6mm} (ii)$_1$ $\varphi(x,y) = \varphi(u, x^{\sharp} y), {}^\forall x,y \in \A$;

\hspace*{6mm} (ii)$_2$ $| \varphi(x,y) | \leq \| x\| \|y \|, {}^\forall x,y \in \A$;

\hspace*{6mm} (ii)$_3$ $ \varphi(x,y) = \varphi(y^*, x^*), {}^\forall x,y \in \A$;

Further, $(\overline{\B}[\| \ \|_1], \sharp_1)$ is standard if and only if 

\hspace*{6mm} (ii)$_4$ $\varphi(x^*, x^{\sharp}) \geq 0, {}^\forall x \in \A$. \\[3mm]
\indent
{\it Proof.} (i) $\Rightarrow$ (ii) We put
\[
\varphi(a,b) = (\Phi(a) | \Phi(b)), \hspace*{5mm} a,b \in \OA[\| \ \|],
\]
where $( \ | \ )$ is the inner product defined by the Hilbertian norm $\| \ \|_1$ on $\overline{\B}[ \| \ \|]$. Then it is easily shown that $\varphi$ is a positive sesquilinear form on $\OA[\| \ \|] \times \OA[\| \ \|]$ satisfying the condition (ii)$_1 \sim $ (ii)$_3$. If $(\overline{\B} [ \| \ \|_1], \sharp_1)$ is standard, then (ii)$_4$ follows from Lemma 2.5. 

(ii) $\Rightarrow$ (i) We put $\N_\varphi= \{ a \in \OA[\| \ \|]; 
\varphi(a,a) = 0\}$. Then $\N_\varphi$ is a subspace of $\OA[\| \ \|]$ and, due to the positivity of $\varphi$, which implies $\varphi(a,b) = \overline{\varphi(b,a)}$ for each $a,b \in \OA[ \| \ \|]$, it follows that the quotient space $\lambda_\varphi(\OA[\| \ \|]) \equiv \OA[\| \ \|]/\N_\varphi= \{ \lambda_\varphi(a) \equiv a + \N_\varphi; a \in \OA[\| \ \|] \}$ is a pre-Hilbert space with inner product: 
$(\lambda_\varphi(a) | \lambda_\varphi(b))_\varphi= \varphi(a,b), a, b \in \OA[\| \ \|]$. We denote by $\| \ \|_\varphi$ the norm defined by the inner product $(\ | \ )_\varphi$ and by $\H_\varphi$ the completion of $\lambda_\varphi(\OA[\| \ \|]) [\| \ \|_\varphi ]$. Since $\A$ is $\| \ \|$-dense in $\OA[\| \ \|]$, it follows that

(ii)$'_2$ $| \varphi(a,b) | \leq \| a \| \| b \|, {}^\forall a,b \in \OA[\| \ \|]$;

(ii)$'_3$ $\varphi(a,b) = \varphi(b^*, a^*),  {}^\forall a,b \in \OA[\| \ \|]$, \\
and since (ii)$'$ and $\| x \| \leq \| x \|_{\sharp}, {}^\forall x \in \A$, it follows that

(ii)$'_1$ $\varphi(x,y) = \varphi(u, x^{\sharp}y), {}^\forall x,y  \in \A_{\sharp}$. \\
By (ii)$'_2$ $\A_\varphi \equiv \lambda_\varphi(\A)$ is a dense subspace of the Hilbert space $\H_\varphi$ and further, it is a $*$-algebra equipped with the multiplication: $\lambda_\varphi(x) \lambda_\varphi(y)= L_{\lambda_\varphi(x)} \lambda_\varphi(y) \equiv \lambda_\varphi(xy)$ and the involution: $\lambda_\varphi(x)^* \equiv \lambda_\varphi(x^*)$. By (ii)$'_3$ the involution $\lambda_\varphi(x) \rightarrow \lambda_\varphi(x)^*$ can be extended to the isometric involution $J_\varphi$ on $\H_\varphi$. By (ii)$'_1$ the linear functional on the C$^*$-algebra $\A_{\sharp} : x \rightarrow \varphi(x,u)$ is positive, and so $\varphi(y^{\sharp}(x^{\sharp}x) y, u) \leq \| x \|^2_{\sharp} \varphi(y,y)$ for each $x,y \in \A$. Hence it follows from (ii)$_1$ that 
\[
\| \lambda_\varphi(x) \lambda_\varphi(y) \|^2_\varphi = \varphi(xy,xy) = \varphi(y^{\sharp} x^{\sharp} xy, u) \leq \| x \|^2_{\sharp} \| \lambda_\varphi (y) \|^2_\varphi
\]
for each $x,y \in \A$, so that $L_{\lambda_\varphi(x)}$ is bounded and $\| L_{\lambda_\varphi(x)} \| \leq \| x \|_{\sharp}$ for each $x \in \A$. 
Thus $\H_\varphi= \overline{\A_\varphi} [\| \ \|_\varphi]$ is a Hilbert quasi $*$-algebra. Further, the map $\lambda_\varphi(x) \rightarrow \lambda_\varphi(x)^{\sharp} \equiv \lambda_\varphi(x^{\sharp})$ is an involution of $\A_\varphi$ and by (ii)$_1$ $L_{\lambda_\varphi(x)}^* = L_{\lambda_\varphi(x)^{\sharp_1}}$ for each $x \in \A$. Hence, $(\overline{\A_\varphi}[\| \ \|_\varphi], \sharp_1)$ is a HCQ$^*$-algebra. We here put $\Phi(a) = \lambda_\varphi(a), a \in \OA[\| \ \|]$. Then it is easily shown that $\Phi$ is a $*$-homomorphism of the strict CQ$^*$-algebra into the HCQ$^*$-algebra $(\overline{\A_\varphi} [\| \ \|_\varphi], \sharp_1)$ satisfying $\Phi(\A_0) = \A_\varphi$, and by (ii)$'_2$ it is contractive. Suppose that $\varphi$ is faithful. 
Then the $*$-representation of the C$^*$-algebra $\A_{\sharp}$ on $\H_\varphi$ defined by $x \rightarrow L_{\lambda_\varphi(x)}, x \in \A_{\sharp}$ is faithful, which implies that $\| L_{\lambda_\varphi(x)} \| = \| x \|_{\sharp}$ for each $x \in \A_{\sharp}$. 
Further, since $\Phi(\A_0)= \A_\varphi$, it follows that $\Phi(\A_{\sharp}) = (\A_\varphi)_{\sharp_1}$ and $\Phi \lceil \A_{\sharp}$ is a $*$-isormorphism of the C$^*$-algebra $\A_{\sharp}$ onto the C$^*$-algebra $(\A_\varphi)_{\sharp_1}$. 
Hence $\Phi$ is a $*$-isomorphism of $(\OA[\| \ \|], \sharp, \| \ \|_{\sharp})$ into $(\overline{\A_\varphi}[\| \ \|_\varphi], \sharp_1)$. 
By Lemma 2.5, the HCQ$^*$-algebra ($\overline{\A_\varphi}[\| \ \|_\varphi], \sharp_1)$ is standard if and only if (ii)$_4$ holds. 
This completes the proof.

\ 

Now the question arises as to whether positive sesquilinear forms as described in (ii) do really exists. The answer is certainly positive due to the existence of standard HCQ$^*$-algebras stated in Theorem 2.9. Indeed, the inner product $<,>$ of a left Hilbert algebra satisfies conditions (ii)$_1$ - (ii)$_4$. 

Furthermore, Theorem 3.2 answers to the main question of this section: any form $\varphi$ over a strict CQ*-algebra $(\OA[\| \ \|], \sharp, \| \ \|_{\sharp})$ with quasi-unit, can be used to construct a HCQ$^*$-algebra where $\OA$ is contractively embedded. 

\ 

{\bf Acknowledgment} We acknowledge the financial support of the Gruppo Nazionale per l'Analisi Funzionale e le Applicazioni del C.N.R. and of the Itaian Ministry of Scientific Research, and of Japan Private School Promotion Foundation. 

\

\end{document}